\documentclass{elsart3p}

\usepackage{graphicx}

\begin{document}

\begin{frontmatter}

\title{Grain - A Java Analysis Framework for Total Data Readout}
\author{P. Rahkila}
\address{Department of Physics, University of Jyv\"askyl\"a, P.O. Box 35 (YFL), FI-40014 University of Jyv\"askyl\"a, Finland}
\ead{panu.rahkila@phys.jyu.fi}

\begin{abstract}
Grain is a data analysis framework developed to be used with the novel Total
Data Readout data acquisition system. In Total Data Readout all the electronics
channels are read out asynchronously in singles mode and each data item is
timestamped. Event building and analysis has to be done entirely in the software
post-processing the data stream. A flexible and efficient event parser and the
accompanying software framework have been written entirely in Java. The design
and implementation of the software are discussed along with experiences gained in
running real-life experiments.
\end{abstract}


\end{frontmatter} 

\section{Introduction}
Nuclear physics experiments are usually instrumented using conventional,
common dead time data acquisition systems, which are triggered by an event
in a pre-defined detector. In decay spectroscopy and Recoil Decay Tagging
(RDT) experiments (in which decay spectroscopy is combined with in-beam
spectroscopy), these systems inherently suffer from dead time losses since
rather wide common gates have to be  used in order to collect all the required
information.  These problems grow worse if either the focal plane count rate
or the common gate width is increased. The former condition often arises from
the fact that the reaction channel under study may form only a minor fraction
of the total counting rate of the implantation detector. In the latter case
one is usually either studying isomeric decays with half-lives of the order of
tens of microseconds or using them as a ``tag'' in Recoil Isomer Tagging (RIT)
experiments. To overcome these problems a novel Total Data Readout (TDR)
method~\cite{2001LA01-TDR} was developed by the GREAT collaboration as part 
of a project to build a highly sensitive tagging spectrometer~\cite{2003PA01-GREAT} .

TDR is a triggerless data acquisition system in which all the electronics
channels operate individually in free running singles mode. All the 
information is read out asyncronously by the front-end electronics consisting 
of gated ADCs and bit-pattern registers. Data items are time-stamped with 10ns
precision using a 100 MHz clock signal, which is distributed throughout the
whole system. The data are subsequently ordered in a so called collate 
and merge software layer, after which the data forms a single time-ordered
stream. Unlike the data emerging  from a conventional data acquisition
system, the data from the TDR collate and merge layer is not structured or
filtered in any way, excluding the time ordering. Temporal and spatial
correlations required to form events out of the raw data stream and the 
filtering to remove unwanted or irrelevant data has to be done entirely in the
software which processes the data stream.

Grain was developed to provide a complete, self-contained, cross platform 
software framework which could be used to analyse the raw TDR data stream. The
main purpose of the software is to provide a tool for the online analysis at 
the RITU separator at the Accelerator Laboratory of the University of
Jyv\"askyl\"a (JYFL), where the TDR system along with the GREAT spectrometer
are currently located, and to facilitate for the subsequent offline analysis
of the experimental data. The GREAT TDR system also includes an event builder
software, TDREB \cite{TDR-EB-WWW}. Grain can be used as a completely
stand-alone system or either in parallel or in series with TDREB.

Grain has been implemented entirely in Java. The portability, clean
object-oriented programming language and the incorporated, rich user interface
and networking libraries were the main motives behind the decision. Java has
had a reputation of being too slow for calculation intensive tasks, such as data
analysis, but in the recent years the arrival of just-in-time (JIT) compilers
have lifted the performance to the same level as native, compiled languages
(see e.g.~\cite{2001-JAVA-BENCH}). Previous reports on the usage
of Java in similar tasks~\cite{2001SW01-JAM,JAS-WWW} were also found to be
mostly positive. The Grain executable is available for download at the
development web page \cite{GRAIN-WWW}.

\section{Stream filtering and event parsing}

\subsection{Stream filtering}
Prior to building the events the TDR data stream must be filtered against
unwanted data, which usually consists of vetoed and piled up signals. In traditional
systems the vetoing and pile-up detection was incorporated into the front-end 
electronics and the data acquisition system normally would not ever see these 
data. In the TDR system the data analysis software is required to perform
these tasks, though the TDR ADCs have a limited hardware-veto capability. For
example, in the current JYFL TDR setup events from the Compton 
suppression shields of the target array are read out as bit-pattern data. Thus,
the data from the target array germanium detectors and their BGO shields must
be correlated pairwise in software in order to perform the suppression. 

Pile-up rejection
is based on the TDR ADCs capability to detect gates arriving at the ADC during
the processing of the previous gate. These data are included in the stream as
separate special data items and thus each channel needs to be self-correlated in
time to find the piled-up data. Vetoed or pile-up data can be either discarded
or marked and included in the events.

\subsection{Event parsing}

\begin{figure*}[!tb]
\includegraphics[width=\textwidth]{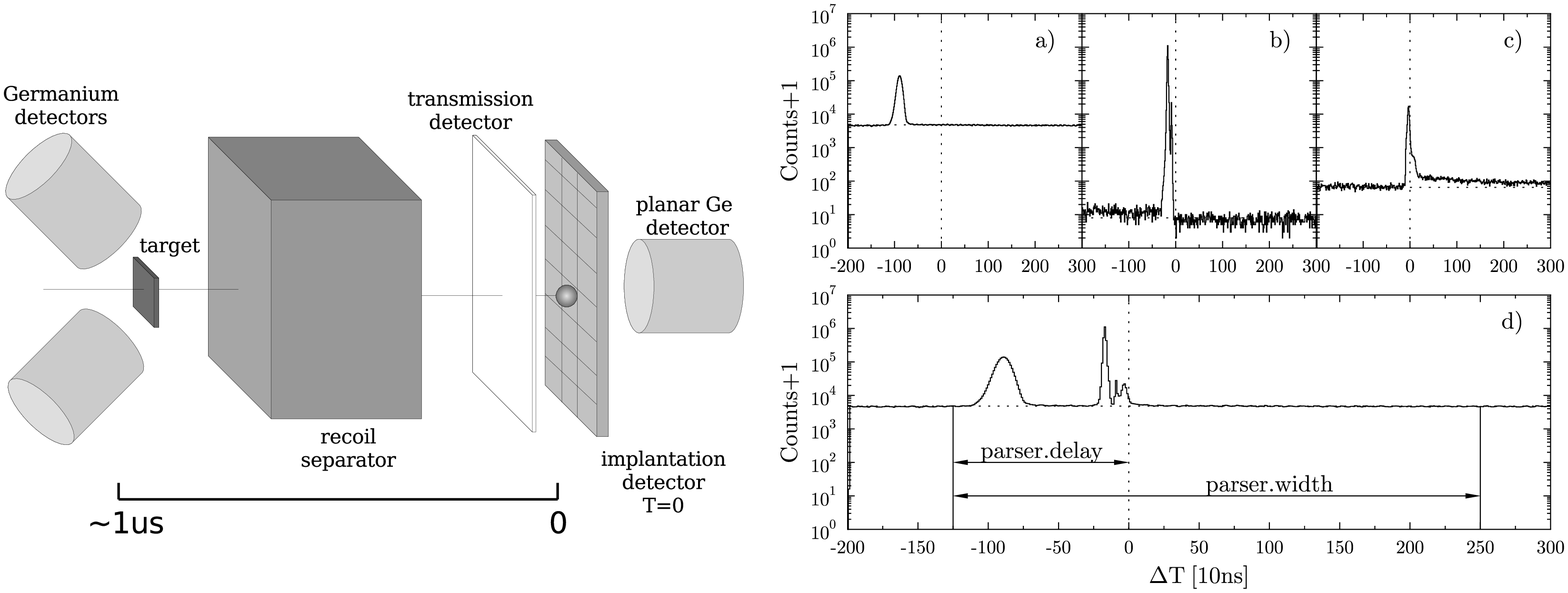}
\caption{Schematic illustration of a typical RDT setup and the time structure of
a typical RDT experiment data stream with respect to any signal in the
implantation detector. Panels a),b) and c) represent the typical response of
some individual detector groups. Panel d shows the summed structure and the main
event builder timing parameters. In panel a) the time spectra of the germanium
array at the target position is shown. Flight time of the recoils through the
separator is $\sim$1$\mu$s. Panel b) shows the timing of a Multiwire
Proportional Counter placed 240 mm upstream from the implantation detector.
Panel d) shows time structure of events in the planar germanium detector placed
next to the implantation detector exhibiting prompt and delayed components.} 
\label{rdttimes}
\end{figure*}

Two types of event parsers have been developed so far. Decay spectroscopy and
tagging experiments require a parser which constructs events in which the
trigger is any signal from the implantation detector. Stand-alone in-beam
experiments require a trigger based on the multiplicity of hits in the
detector array. In both cases time domain correlations were selected as the first stage
of the event builder strategy. This was mainly done in order to ensure the maximum 
throughput of the system as the conditions used require only a single dynamic
parameter, the time stamp and a static definition of which data acquisition
channels constitute the triggering detector group.

The decay/RDT event parser is almost entirely based on the time structure
of the stream. A typical time structure of the stream, with respect to any
signal in the implantation detector, taken from a tagging experiment at RITU is presented in figure
\ref{rdttimes}. The individual components forming the structure can be roughly
divided into three groups depending on the placement and role of the detector
groups: a) preceding, b) prompt and c) delayed events. Typical examples of
these are presented in the upper panels of figure \ref{rdttimes}. 

In decay or RDT experiments events can be simply defined at the first stage as a
time slice of the stream, which is triggered by any datum from a predefined
group of ADC channels. As the data is buffered in time order, it is possible to
easily extend the slice to cover also data in the past and in the future with 
respect to the triggering data. The parameters needed to construct the slice 
are the address of the triggering channel, offset of the slice (delay)
and the extent of the slice (width). By varying these parameters the parser can
be configured for different types of requirements of RDT, RIT or
decay spectroscopy experiments.

\begin{figure}[b]
\includegraphics[width=0.45\textwidth]{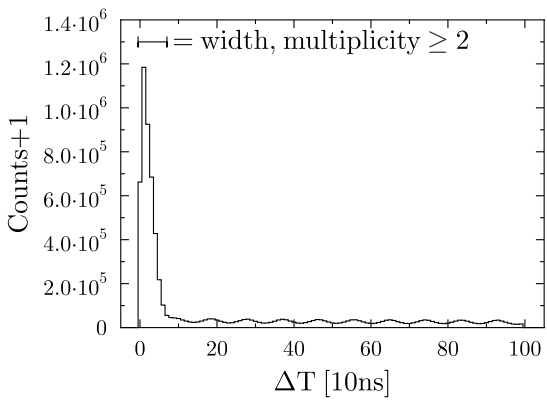}
\caption{Time structure of the TDR stream from a stand-alone $\gamma$-ray
experiment. Data is histogrammed if more than one hit is registered in
1$\mu$s window. In normal running conditions a 70 ns window would be used, as
indicated in the figure. Oscillations in the background are caused by the
structure of the beam produced by the JYFL K130 cyclotron.}
\label{ggtimes}
\end{figure}
Pure in-beam experiments usually use a hit-multiplicity trigger; a certain
number of coincident hits is required in a defined time window. In the case of
TDR the event parsing is rather straight forward. As the data is already time
ordered and filtered, and can be easily buffered in memory, one can simply
count the number of hits over a given period after each individual hit.
The input parameters required are the width of the coincidence window, the
set of channels for which the multiplicity is calculated for and the
minimum required multiplicity. Figure \ref{ggtimes} shows a typical time
structure of the TDR data stream in a stand-alone $\gamma$-ray experiment. 

Both event parsers have been implemented around a ring-buffer, which holds the
data objects. Access to the data preceding and following a certain data item
can be done by iterations in the buffer. On insertion and removal the data item
is checked whether it is a triggering item, a vetoing item or a piling-up item.
In any of these cases corresponding data is searched in the buffer and flagged
if found. All the data is checked on removal whether it is flagged triggered
and if so, dispatched to the next level of event parser.
\begin{figure}[b]
\includegraphics[width=0.45\textwidth]{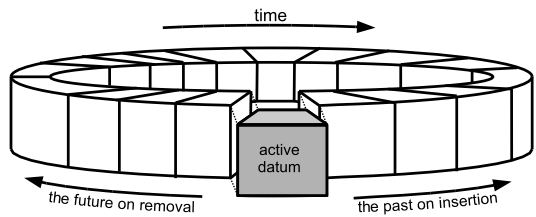}
\caption{Event search is performed in a ring-buffer, which provides effient
access to data preceding and following any data item.}
\label{buffer}
\end{figure}

\begin{figure}[t]
\includegraphics[width=0.45\textwidth]{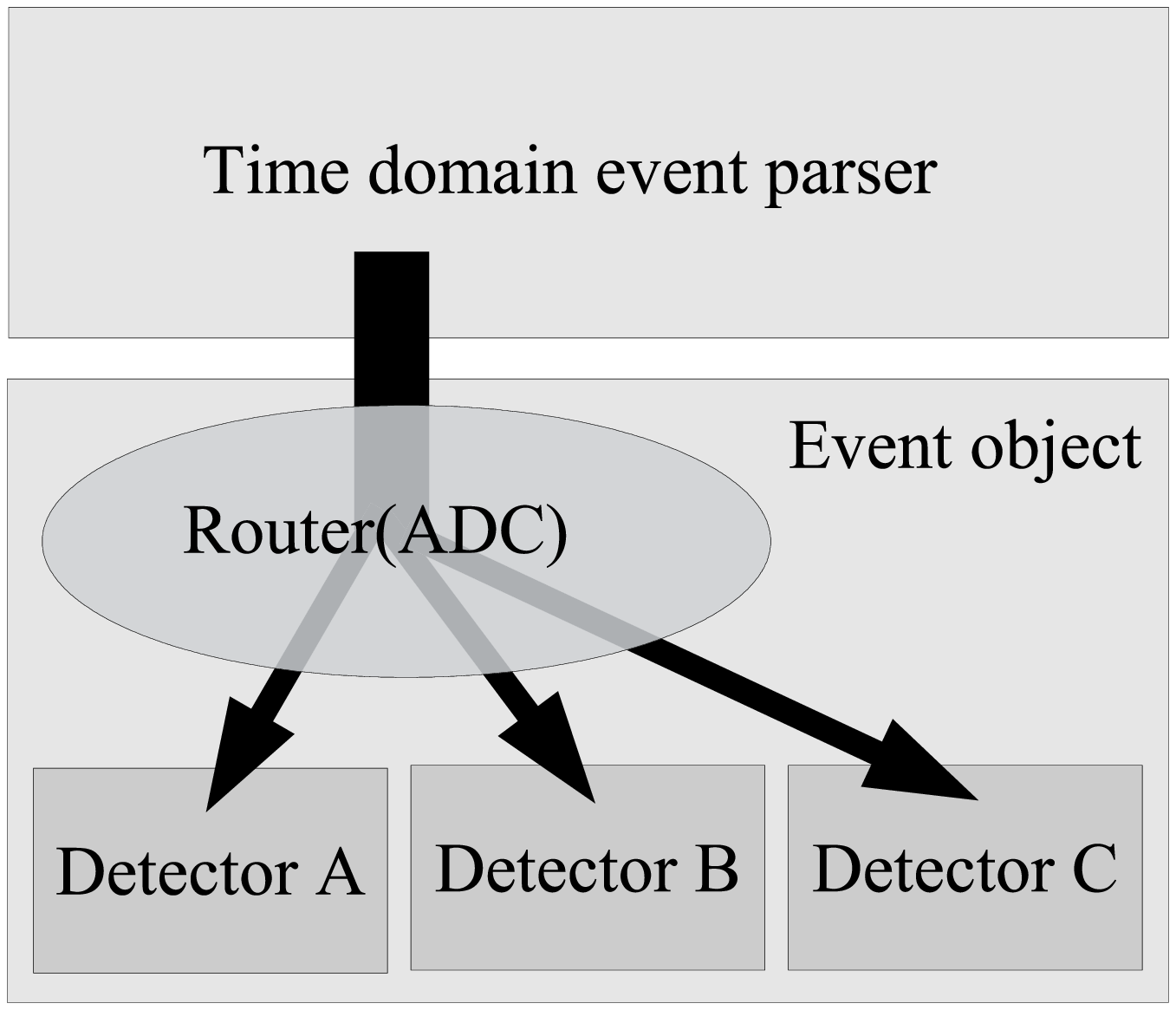}
\caption{Data is routed to the detector objects within the event object based
on the origin of the data.}
\label{route}
\end{figure}

Once the group of data forming an event has been identified, the internal event
structure needs to be assigned. At the second stage of event building the data
forming the event slice are fed into an event object. The data items are routed
into sub-objects describing different detectors according to the source of the
data i.e. the ADC or bit-pattern unit channel number. The routing table is
defined in the sub-detector objects at the implementation time. A schematic
drawing of the routing is shown in figure \ref{route}. Several different event
types have been predefined and users select the type which is appropriate
for their analysis from the user interface.

\section{Framework implementation}
\begin{figure}[b]
\includegraphics[width=0.45\textwidth]{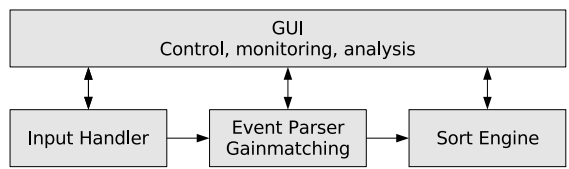}
\caption{Schematic of the data processing framework. Each data processing
sub-task runs on a separate thread which are interconnected by FIFOs and
supervised by the master thread running the GUI.}
\label{schemdes}
\end{figure}

The general design of the analysis framework is shown in figure \ref{schemdes}.
In order to benefit from modern computer hardware with multiple processor
cores available on most machines, the data processing framework was designed to
be multithreaded. Users interact with the graphical user interface (GUI) running
the master thread and providing interactive analysis functions as well as
serving as the control thread for data sorting jobs. The GUI thread starts the
other threads at the beginning of each data sorting job. The data is relayed
from thread to thread using first-in, first-out buffers (FIFOs).

\subsection{Sort Engine}
The sort engine uses the Java dynamical class loading capability. Grain
provides abstract (skeleton) sorter classes which the user needs to implement
and which provide access to the event data. Users can thus write their own
data reduction routines in Java using all the features of the language as long
as this inheritance relationship is fulfilled. Compiled classes
can be loaded into the Java Virtual Machine (JVM) dynamically at
runtime. Histogramming and other basic analysis services are provided via
JAIDA~\cite{JAIDA-WWW}, the Java implementation of the AIDA (Abstract
Interfaces for Data Analysis) definition~\cite{AIDA-WWW}. A new binner had to
be added to the JAIDA histogrammer since rather large multidimensional
histograms are required in nuclear physics analysis. The histograms and
n-tuples created in the sort engine are available through the GUI at runtime.

\subsection{Correlation Framework}
During the last decade the RDT technique
\cite{NSR1986SI19-HG180-RDT,NSR1995PA01-I109-RDT} has been widely used in the studies of the
structure of neutron deficient nuclei and super-heavy nuclei (see e.g. review
articles~\cite{NSR2001JU09-HGPBPO-REV,NSR2004HE04-SHE-REV}). The technique has
usually been applied in set-ups where one detector system is used to observe
prompt radiation at the target  position of a recoil separator, while the other is
located at the focal plane of the separator detecting the arrival of reaction
residues (recoils) and their subsequent decays. In essence, RDT is a three
step process. First a delayed coincidence is used to associate the prompt
radiation with the recoils implanted in a position sensitive detector at the
focal plane. In the second step a spatial and temporal correlation of the
recoil and the subsequent, often discrete, decay is used to establish the
identity of the recoil. Finally, this information can be combined to form the
unambiguous identification of the source of the prompt radiation. Similar correlation
tasks are used also in pure decay spectroscopy. A large amount of information
per event needs to be stored often for several hours in order to perform these
correlations.

The Grain correlation framework is based on the discrete position sensitivity
provided by the double-sided silicon strip detectors used in GREAT and the fact
that all the event information is already encapsulated in the event object. The
framework consists of a container object which provides a time ordered,
time constrained stack of event objects per implantation detector pixel and
routines to insert an event into the container and to retrieve the history of
any given pixel based on the current event. This framework simplifies
correlation analysis greatly as the user does not need to implement book
keeping and memory management and they are handled in a consistent manner for
all the users.

\subsection{User Interface and Analysis Functions}
\begin{figure*}
\includegraphics[width=\textwidth]{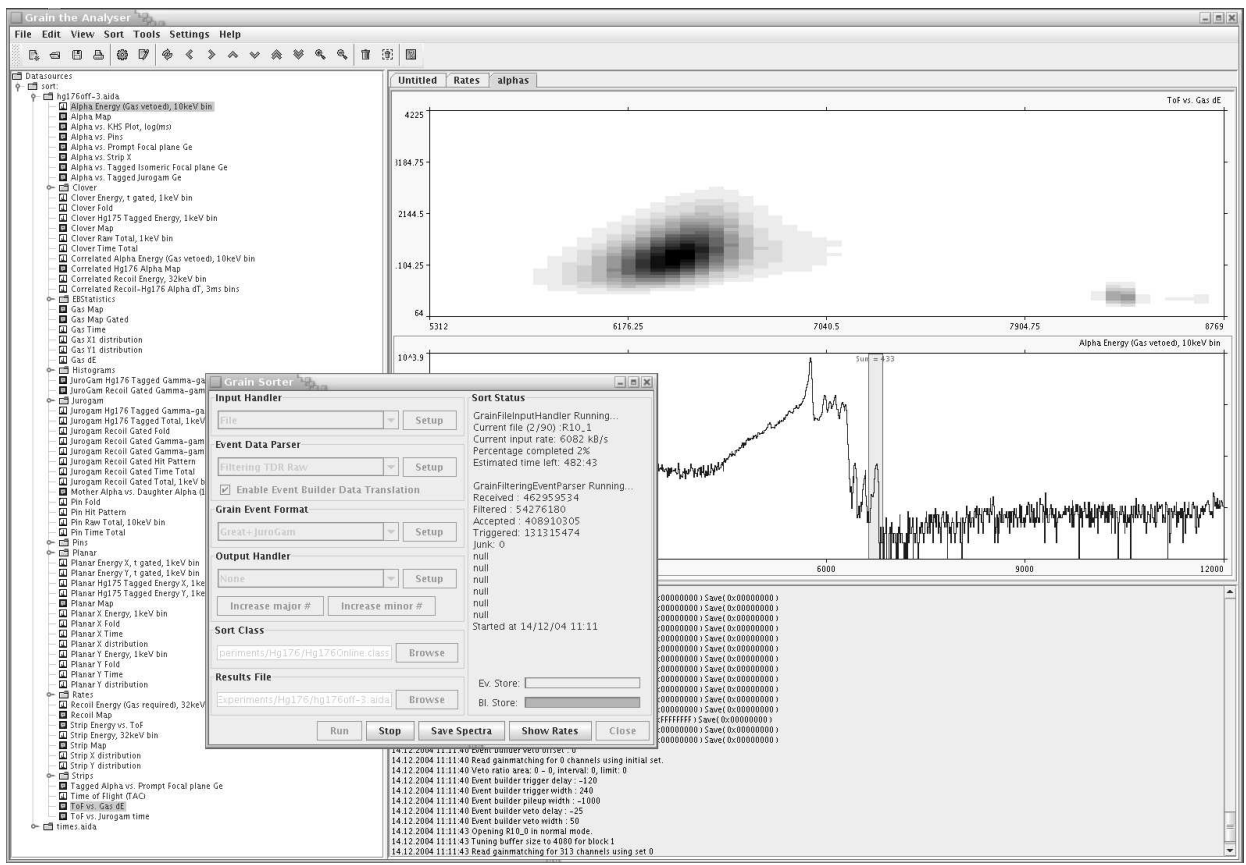}
\caption{Screenshot of the Grain GUI. Main window displaying two- and
one-dimensional histograms, histogram tree and logger window along with the
sort control window are shown.}
\label{sshot}
\end{figure*}

The Grain graphical user interface has been implemented using standard the Java
Swing toolkit and Java2D graphics (see fig.~\ref{sshot}). A standard GUI design, with
menus and toolbars, was selected as it is already familiar to most users.
Histograms can be browsed with a tree widget and displayed on the main panel,
several at a time if required. Information about the progress of the 
analysis jobs and the results for the interactive analysis are displayed in
the logger window. Standard zooming and scrolling functions are provided for
both one- and two-dimensional histograms. Peak-area integration and fitting of
gaussian peak-shapes and exponential decay-curves are provided for
one-dimensional histograms. Two-dimensional histograms (matrices) can be
sliced either on the standard GUI or on a separate widget geared towards
coincidence analysis. N-tuples can also be used in the interactive analysis
through AIDA evaluators and filters.
Histograms can be exported to ascii and Radware~\cite{RADWARE-WWW} formats. 
The ascii format can also be imported along with ROOT histograms through the
hep.io.root package~\cite{ROOTIO-WWW}. The spectrum view can be printed using the printing
system provided by the operating system or exported to a variety of formats.
The AIDA XML data format used by Grain through JAIDA libraries to store
histograms and n-tuples is an open standard. Several other AIDA-compliant
tools~\cite{JAS-WWW,OPENSCIENTIST-WWW,PI-WWW,PAIDA-WWW} can be used to read, view and
analyse the files instead of Grain if so required.

\section{Performance}
The sorting performance has been analysed in two ways. First, only
simple through-put tests were run on a modern computer with a dual-core AMD
processor running 64-bit linux operating system and 64-bit Java version 1.6
from Sun Microsystems. Later, the performance of different parts of the data
sorting chain have been analysed using the Netbeans Java
Profiler~\cite{NETBEANS-PROFILER-WWW}.

To demonstrate the performance of the sorting a typical RDT experiment was
selected as a test case. A heavy ion fusion evaporation reaction
$^{36}$Ar~+~$^{144}$Sm~$\rightarrow$~$^{180}$Hg$^*$ was used to produce light
Hg isotopes. In optimum operating conditions the total counting rate of the 
detectors was about 400 kHz, mainly from the target array germanium detectors, 
corresponding to a data rate of about 3.2 MB/s from the TDR. Part of
the data was written to disk without any prefiltering, and later analysed offline. The 
results are shown in Table \ref{perf}. In early experiments the histogramming 
of raw data was noticed to have a serious impact on the sorting performance
on-line. This is clearly reflected in the results. Histogramming of the raw data
is also performed in the TDR DAQ, so it can be safely turned off if the
performance degradation is too high. As can be seen, the throughput without
raw histogramming is over an order of magnitude higher than a typical
RDT experiment currently requires and close to that for stand-alone experiments.
In decay experiments data rates are always much lower as the target area 
detectors are not used.

\begin{table}[ht]
\caption{Throughput of the Grain sorter. See text for details.}
\begin{tabular*}{0.45\textwidth}{@{\extracolsep{\fill}} c c c c }
\hline
Trigger & with raw & w.o. raw \\
& histogramming & histogramming \\
\hline
RDT & 19 MB/s & 44 MB/s \\
$\gamma\gamma$ & 13 MB/s & 22 MB/s\\
$\gamma\gamma\gamma$ & 17 MB/s & 28 MB/s \\
\hline
\end{tabular*}
\label{perf}
\end{table}

The event parser has been found to be the bottleneck in the sorting performance
by using the Java profiler. About 65\% of the execution time is spent in the
event parser thread, out of which about a half is spent in the actual event
search in the ring-buffer. This bottleneck is partly alleviated by the
multithreading as the parser utilises a single processor core and the other
parts of the framework run in the other available cores. 

\section{Conclusions}
Analysis of the triggerless, TDR generated data has been implemented in a
flexible, efficient manner. Grain has been used as an on-line analysis tool in
over 50 experiments since 2002, catering for very different experiments ranging
from decay spectroscopy of very heavy elements~\cite{NSR2006HE19-NO-NATURE} to
RDT studies in the A$\sim$100 region~\cite{NSR2007SA36-XE110}and the
development of the novel $\beta$-tagging technique~\cite{NSR2006ST14-BTAG-NIM}.
In vast majority of cases Grain has also been the main tool in offline
analysis.

The use of Java language and platform has been a major contributor to the
success of the framework. Platform independence has granted simple installation
and operation on the three current major personal computer operation systems,
making it easy for users to deploy the software where required. Java language
and the use of object oriented techniques has not only simplified development
of the framework itself, but has simplified the users task of sort code writing,
especially when complicated correlation schemes have to be used. 

\section*{Acknowledgments}
I would like to thank all the people who have used the software and contributed
through bug reports and suggestions. This work has been supported by the EU
5th Framework Programme ``Improving Human Potential - Access to Research
Infrastructure''. Contract No. HPRI-CT-1999-00044 and the EU 6th Framework
programme ``Integrating Infrastructure Initiative - Transnational Access'',
Contract Number: 506065 (EURONS) and by the Academy of Finland under the
Finnish Centre of Excellence Programmes 2000-2005 (Nuclear and Condensed
Matter Physics Programme at JYFL) and 2006-2011 (Nuclear and Accelerator Based
Physics Programme at JYFL).

\end{document}